\documentclass{article}
\usepackage[dvips]{graphicx}
\begin{document}
\title{Sidebranch Structures of Dendritic Patterns in a Coupled Map Lattice Model}
\author{Hidetsugu Sakaguchi$^{a}$ and Masako Ohtaki$^{b}$\\
$^{a}$Department of Applied Science for Electronics and Materials, \\
Interdisciplinary Graduate School of Engineering Sciences, Kyushu
University, \\
Kasuga, Fukuoka 816-8580, Japan\\
$^{b}$ Advanced Material Institute, Fukuoka University,\\
Nanakuma, Fukuoka 8-19-1,Japan}
\maketitle
\begin{abstract}
Dendrites with developed sidebranches are numerically studied with a coupled map lattice model. The competitive dynamics among sidebranches determines the shape of the envelope.
The envelope has a parabolic shape near the tip of the dendrite and the envelope angle with respect to the main branch increases up to 45$^{\circ}$ finally. In an intermediate region, the envelope grows roughly in a power law, however, the exponent increases gradually as a function of the distance from the tip. The competitive dynamics among many branches is also observed in a unidirectional growth from a linear seed, and it is compared with the competitive dynamics of sidebranches. 
\end{abstract}

\section{Introduction}
Pattern formations in diffusion fields have been studied as a
typical nonlinear and nonequilibrium problem.\cite{rf:1,rf:2}
The dendrite is a typical pattern observed in crystal growth.
The tip velocity and the shape near the tip region have been intensively studied. The growth law of the dendrite is expressed as $v\rho^2=const$, where $v$ is the tip velocity and $\rho$ is the radius of curvature of the parabolic tip.\cite{rf:3,rf:4} The parabolic structure of the dendritic tip is linearly stable and the formation of sidebranches is considered to be due to some noise effects.\cite{rf:5} Perturbations  are expected to grow as  $\exp(\gamma x^{1/4})$, where $x$ is the distance from the dendritic tip,\cite{rf:6} and they grow into sidebranches. Well-developed sidebranches can be easily observed in experiments of succinonitrile and NH$_4$Cl. There are some quantitative investigations about the well-developed sidebranches. Huang and Glicksman investigated the development of sidebranches in an experiment of succinonitrile.\cite{rf:7} They found that the sidebranching spacing, which is the interval between neighboring active sidebranches, grows in a power law  of exponent 1.3 as a function of distance from the dendritic tip.  
They found the angle of the envelope with respect to the main branch increases as a function of the distance from the tip until 41$^{\circ}$. They expected that the angle increases up to 45$^{\circ}$ finally, where the sidebranches grow with the same velocity as the main branch. 
Li and Beckermann analyzed the shape of the envelope of the sidebranches for succinonitrile on the microgravity flight and found that the envelope grows with a power law of exponent 0.859 as a function of the distance from the dendritic tip.\cite{rf:8}  
The exponent is less than 1, and it implies that the envelope angle decreases as a function of distance from the tip, which is different from the observation by Huang and Glicksman. Recently, Honda and Honjo are investigating the shape of the sidebranches of kinetic dendrites in viscous fingering on the plate with many grooves of square symmetry.\cite{rf:9}  Theoretically, well-developed sidebranches have not been studied in detail. We will study well-developed sidebranches using a coupled map lattice model.  

We proposed a coupled map lattice model as a simple simulation method for crystal growth and succeeded in generating various patterns such as DLA, DBM and dendrites.\cite{rf:10} 
The dendrite with anisotropic surface tension obeys the growth law $v\rho^2=const$ also in our coupled map lattice, when the undercooling is sufficiently small.\cite{rf:11} The model is a deterministic model similar to the phase field model.\cite{rf:12}  Our coupled map lattice models involves the kinetic effect and we have not succeeded in removing the kinetic effect as in the phase-field model by Karma and Rappel.\cite{rf:12} However, our coupled map model is much simpler and numerically preferable for a large scale simulation even at small undercooling such as $\Delta\sim 0.1$.  Previously, different types of coupled map lattice models were proposed by Kaneko and Kessler et al.\cite{rf:13,rf:14} Their coupled map lattice models may be interpreted as direct discretization of the phase-field models, but they have not reported dendritic growth. Our model is a simpler model, in that only the linear diffusion process is assumed in lattice points other than interface sites. Packard and Li-Godenfeld proposed cellular automaton models for the crystal growth.\cite{rf:15,rf:16} In their cellular automaton models, the order parameter takes 0 (liquid) or 1 (solid). In our coupled map lattice model, the order parameter takes a continuous value between 0 and 1 at the interface sites. The interface grows therefore more smoothly than the cellular automaton models. Our coupled map lattice model is considered to be a preferable model compared to the previous coupled map lattice models and cellular automaton models. We will perform numerical simulation of dendrites using the coupled map lattice model to study the growth law of the envelope.   
\begin{figure}[htb]
\begin{center}
\includegraphics[width=15cm]{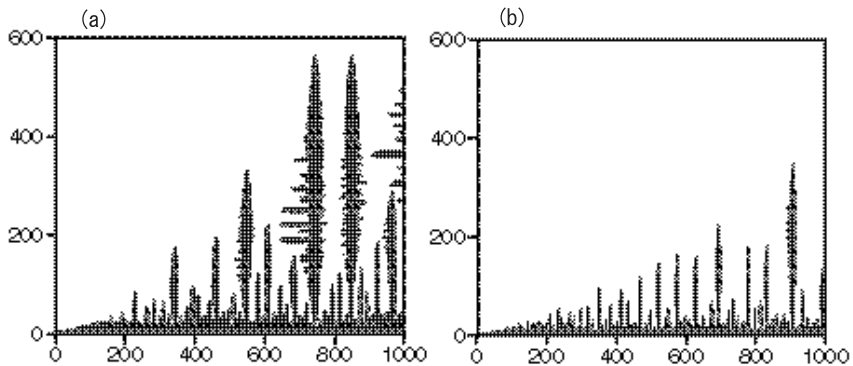}
\end{center}
\caption{Dendritic patterns in a coupled map lattice for (a) $c_2=0.3$ and (b) 0.7. The dendritic patterns are plotted as the tip position locates at the origin.}
\label{fig:1}
\end{figure}

\begin{figure}[htb]
\begin{center}
\includegraphics[width=13cm]{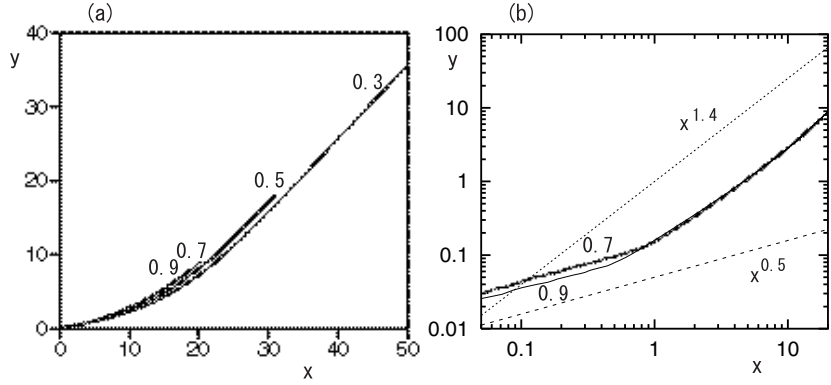}
\end{center}
\caption{(a) Envelope profiles for $c_2=0.3,0.5,0.7$ and 0.9, which are plotted with the coordinates  scaled with the diffusion length $2D/v$.
The numbers in the figure denote the values of $c_2$
(b) Double-logarithmic plots of the envelopes for $c_2=0.7$ and 0.9.
Two linear lines denote power laws with exponents 0.5 and 1.4.}
\label{fig:2}
\end{figure}

\begin{figure}[htb]
\begin{center}
\includegraphics[width=8cm]{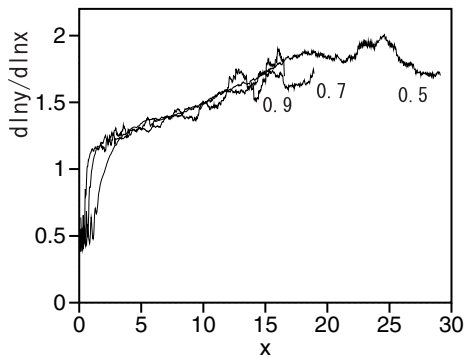}
\end{center}
\caption{The derivative $d\ln y/d\ln x$ as a function of the scaled distance 
from the tip $x$ for $c_2=0.5,0.7$ and 0.9.}
\label{fig:3}
\end{figure}

\begin{figure}[htb]
\begin{center}
\includegraphics[width=7cm]{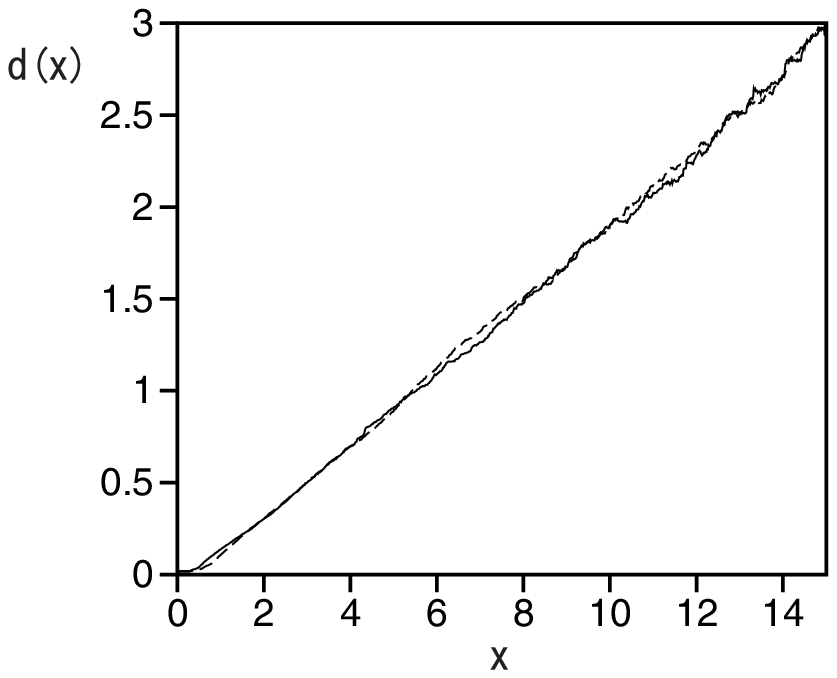}
\end{center}
\caption{Sidebranch spacing $d(x)$ as a function of the scaled distance from the tip $x$ for $c_2=0.9$ (solid curve) and 0.7 (dashed curve).}
\label{fig:4}
\end{figure}

\section{Sidebranch structure of dendrites in a coupled map lattice}
We consider melt growth of crystal with four-fold rotational symmetry. 
Our model consists of two steps of time evolution on a square lattice.  
The four-fold rotational symmetry of crystal is involved in performing the simulation on the square lattice.  If the triangular lattice is used, a pattern like a snowflake with the six-fold rotational symmetry grows as shown in the previous paper.\cite{rf:11} 
The first step is a diffusion process and the second step is a growth process at the interface.
The diffusion process of the latent heat released at the interface is expressed as  
\begin{eqnarray}
u_{n+1}^{\prime}(i,j)& =&u_{n}(i,j)+D\{u_{n}(i+1,j)+u_{n}(i-1,j)+\nonumber\\
& &u_{n}(i,j+1)+u_{n}(i,j-1)-4u_{n}(i,j)\},
\end{eqnarray}
where $u,u^{\prime}$ denotes a dimensionless temperature, $n$ is the step number, $(i,j)$ denotes the lattice point and \(D\) is a diffusion constant.
 This process is interpreted as the simplest discretization of the Laplacian. The second step is a growth process at the interface.  The order parameter is expressed with the variable \(x(i,j)\), where \(x(i,j)\) is 0 in the liquid phase and 1 in the solid phase. 
The order parameter $x(i,j)$ changes only at the interface sites between the liquid sites and the solid sites.   
The growth rule is written as
\begin{eqnarray}
x_{n+1}(i,j) & = & x_{n}(i,j)+c_{1}(1-u_{n+1}^{\prime}(i,j)),\nonumber\\
u_{n+1}(i,j) & = & u_{n+1}^{\prime}(i,j)+c_{2}(1-u_{n+1}^{\prime}(i,j)),
\end{eqnarray}
where the melting temperature is assumed to be 1. 
There are two parameters $c_1$ and $c_2$ in our model, and the ratio $c_2/c_1$ means the latent heat. This growth process is done only at the interface sites, and $x_{n+1},u_{n+1}$ for the other sites keep the values $x_{n},u_{n+1}^{\prime}$ at the previous step. When $x(i,j)$ goes over a threshold $x_c=1$, the interface site changes into a solid site. The diffusion process (1) and the growth process (2) at the interface are performed alternately. 
The initial temperature is assumed to be $u(i,j)=u_0=0$. Then, the normalized undercooling $\Delta$ is given by $\Delta=(1-u_0)/(c_2/c_1)=c_1/c_2$.
We have set a seed of crystal at the origin as an initial condition. To save computation time, we have calculated only the upper half of the crystal assuming the mirror symmetry.  The system size is $1200\times 1200$ in most simulations. The parameter $c_1$ is fixed to be 0.1 and the diffusion constant is $D=0.2$. The surface tension effect is neglected in this paper. (A stable parabolic structure is obtained if the term for the surface tension is included in this model as shown in ref.11.) Our dendrite is therefore a kinetic dendrite, since the growth direction is determined by the kinetic anisotropy.\cite{rf:17} Our simulation may be closely related to the experiment by Honda and Honjo, since they investigate  kinetic dendrites in two dimensions.  
 
Figure 1 displays snapshot patterns of dendrites for (a)  $c_2=0.3$ and (b) $c_2=0.7$. To include small noise effect, we have assumed that the threshold  $x_c$ at the tip site (only one site) takes a random value between 0.99 and 1.01. 
Sidebranches are well developed in the region far from the tip. The sidebranches which locate in $x>1000$ are not shown.  The sidebranches are more developed for $c_2=0.3$ than for $c_2=0.7$, since the undercooling is large. We have calculated the envelope line for each snapshot pattern and constructed an average curve of the envelope for $c_2=0.3,0.5,0.7$ and 0.9. In our numerical simulation, an envelope of sidebranches is defined as the line which connects the tip positions of active sidebranches. A sidebranch is judged as active if it is larger than the neighboring active sidebranch locating in the left of itself (with smaller $i$). As a result of our definition, the envelope is a monotonously increasing function. A relevant length scale of our model is the diffusion length $l=2D/v$. The tip velocities are numerically obtained respectively as 0.0058,0.0081,0.0124 and 0.0227 for $c_2=0.9,0.7,0.5$ and 0.3. The velocities do not obey the law $v\sim \Delta^4$ like the surface tension dendrite in two dimensions, but seem to obey $v\sim \Delta^{1.5}$ for the kinetic dendrites in our model. Figure 2(a) displays the averaged envelope, which is plotted with the scaled length unit $x=i/l,y=j/l$. (We have performed numerical simulation with larger size $1500\times 1500$ only for $c_2=0.9$, since the development of sidebranches is slow at the parameter.) 
The envelope for $c_2=0.3$ can be approximated as $y=x-14.5$ for large $x$. That means that the angle of the envelope with respect to the $x$-direction approaches 45$^{\circ}$. The envelopes overlap with each other rather well, especially, the envelope for $c_2=0.9$ overlaps with that for $c_2=0.7$ very well. There may exist a universal scaling function for the envelope function for sufficiently small undercooling. Figure 2(b) displays logarithmic plots of the envelope functions for $c_2=0.7$ and 0.9. For small $x$, the envelope is approximated as $x^{1/2}$, where the sidebranches are not well developed and the parabolic structure of the tip region is maintained. For relatively larger $x$, the envelope, roughly speaking, obeys a power law with exponent about 1.4. However, the envelope in the logarithmic scale has a tendency to curve upwards as $x$ is increased. The derivative $d\ln y/d\ln x$ is calculated and displayed in Fig.~3 for $c_2=0.5,0.7$ and 0.9. The derivative is about 0.5 for small $x$, which corresponds to the parabolic shape, however, it increases from 1.2 to 1.9 in the intermediate region as $x$ is increased.  This implies that the envelope does not obey a simple power law. We have also calculated the spacing between neighboring active sidebranches and obtained the average curve of the spacing between the active sidebranches as a function of the distance from the tip position. Figure 4 displays the average spacing $d(x)$ for $c_2=0.7$ and 0.9 as a function of the scaled distance $x$. The two curves again overlap very well. The sidebranch spacing is almost a linear function of $x$. By the fitting to a power law $d(x)\sim x^{\beta}$, the exponent $\beta$ is evaluated to be about 1.15. 
The sidebranch spacing increases with the distance from the tip position because of the competition among sidebranches. Smaller sidebranches stop to grow and the survived sidebranches become larger. Competitive processes between survived large sidebranches go on further, until the spacing between the sidebranches becomes sufficiently larger than the diffusion length. 

\begin{figure}[htb]
\begin{center}
\includegraphics[width=18cm]{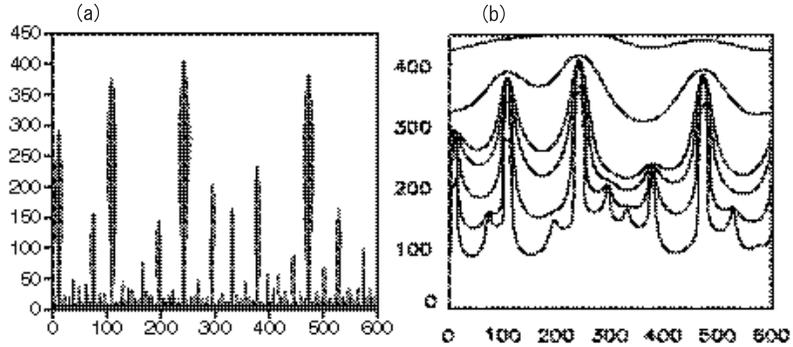}
\end{center}
\caption{(a) Snapshot pattern of dendritic branches for the unidirectional growth from a linear seed for $c_2=0.7$. (b) Contour curves of $u=0.1,0.5,0.9,0.95,0.99$ and $0.999$ for the temperature field $u$ at the same time as (a).}
\label{fig:5}
\end{figure}

\begin{figure}[htb]
\begin{center}
\includegraphics[width=13cm]{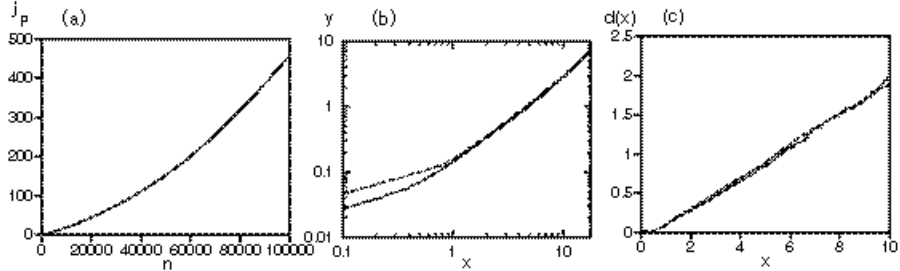}
\end{center}
\caption{(a) Time evolution of the tip position $j_p$ of the largest branch for $c_2=0.7$. (b) Comparison of the time evolution of the tip position of the largest branch for the unidirectional growth from a linear seed (solid curve) with the envelope shown in Fig.~2 for $c_2=0.7$ (dashed curve). They overlap very well.(c) Comparison of the time evolution of the spacing between the neighboring large branches for the unidirectional growth (solid curve) with the spacing between the active sidebranches shown in Fig.~4 for $c_2=0.7$ (dashed curve).}
\label{fig:6}
\end{figure}

\begin{figure}[htb]
\begin{center}
\includegraphics[width=13cm]{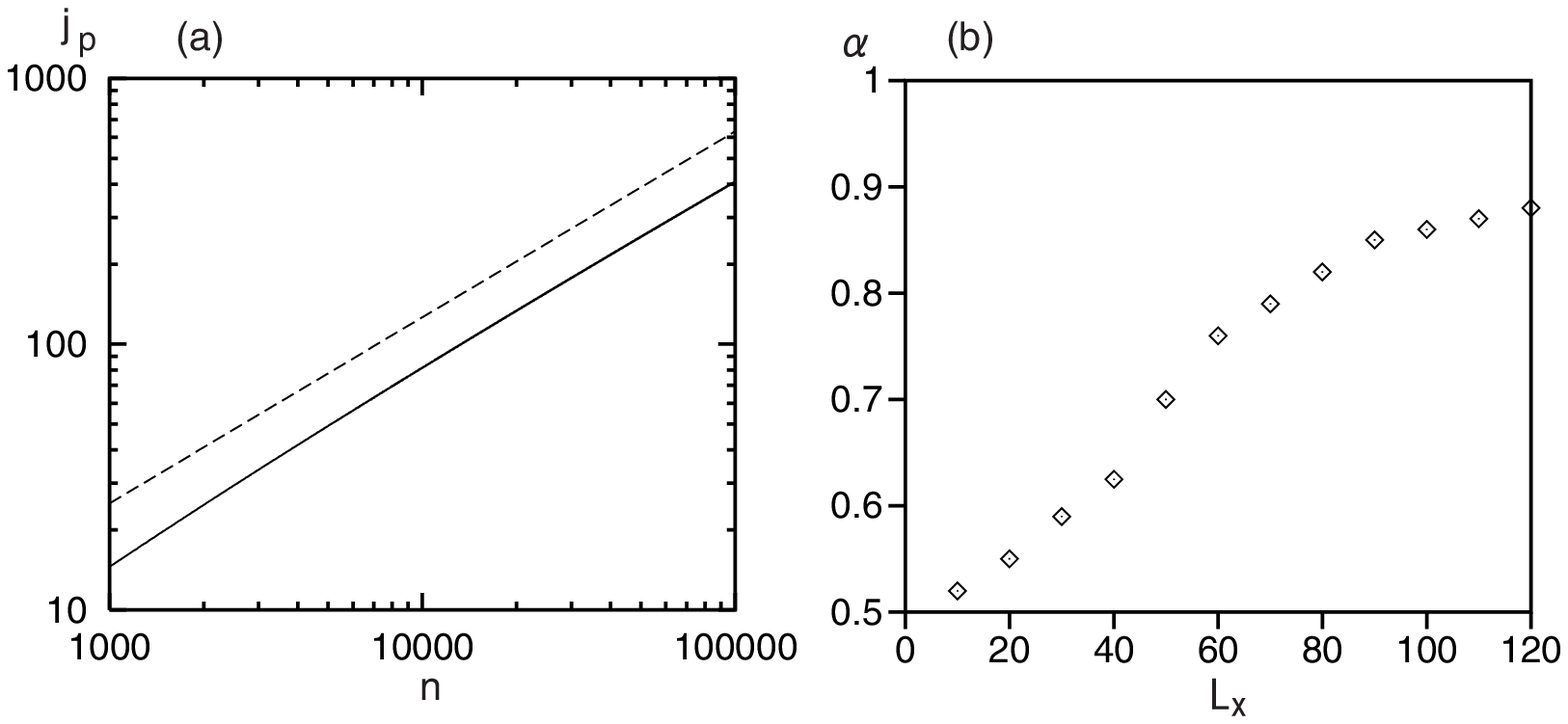}
\end{center}
\caption{(a) Time evolution of the tip position of one branch in a rectangular box with size $50\times 700$ for $c_2=0.7$. The linear line denotes the power law with exponent 0.7. (b) Numerically evaluated exponent $\alpha$ of the power-law growth as a function of the width $L_x$ of the rectangular box.
}
\label{fig:7}
\end{figure}

\section{Competitive dynamics among dendritic branches in the unidirectional growth from a linear seed}
We can see the competitive growth of dendritic sidebranches more explicitly using another type simulation, which is the crystal growth starting from a linear seed. We have performed a numerical simulation in a square box of $L_x\times L_y=600\times 600$. We have imposed  periodic boundary conditions in the $i$ direction such as $u(L_x,j)=u(1,j), u(0,j)=u(L_x-1,j)$ at $i=L_x$ and 0, and fixed boundary conditions such as $u(i,L_y)=0$ at $j=L_y$ and no-flux boundary conditions $u(i,0)=u(i,1)$ at $j=0$. As an initial condition for a linear seed, we assumed that  $x(i,1)=1,u(i,1)=1$, $x(i,2)=0$, and $u(i,2)$ took a random number between 0 and 0.1. The initial value of $x(i,j)$ and $u(i,j)$ for $j\ge 3$ are all 0. 
Figure 5(a) is a snapshot pattern of $x(i,j)$ at $c_2=0.7$.  Many small dendritic branches are created initially, but the number of growing branches decreases in time by the competitive interaction between the neighboring branches.  The velocities of delayed branches decrease and finally become almost zero.  Large branches seen in Fig.~5(a) are survived branches for the competitive process. Figure 5(b) displays contour lines of $u=0.1,0.5,0.9,0.95,0.99$ and $0.999$. 
The large neighboring branches interact competitively through the diffusion field $u$. The temperature $u$ is almost 1 in the regions between the delayed branches.
 
Figure 6(a) displays the time evolution of the tip postition $j_p$ of the largest branch as a function of the step number $n$. The ensemble average with respect to 20 different initial conditions is taken. 
The form of this time evolution is similar to the envelope of sidebranches shown in Fig.~2. We have compared the two curves by rescaling the coordinates. 
The step number $n$ can be converted into a length by $i=vn$ where $v$ is the tip velocity of the main stem and the scaled coordinates $x=i/(2D/v)$ and $y=j/(2D/v)$ are used in Fig.~6(b). Figure 6(b) displays the relation of $y=0.67 j_p/(2Dv)$ and $x=nv/(2D/v)$ (solid curve) and it is compared with the envelope of sidebranches  at $c_2=0.7$ (dashed curve).  The two plots overlap very well, although the numerical factor 0.67 is suitably chosen.   We have also calculated the time evolution of the average interval $\tilde{d}(n)$ between active branches for the crystal growth starting from a linear seed. Here, the active branches at step $n$ are chosen as the branches whose tip positions are larger than $0.7j_p$.  The average interval between active branches increases almost in proportion to $n$, which corresponds to the linear growth of the sidebranch spacing shown in Fig.~4. Figure 6(c) compares the time evolution of the rescaled quantity $d(x)=0.8\tilde{d}(n)/(2D/v)$ as a function of $x=nv/(2D/v)$ (solid curve) with the average spacing between active sidebranches shown in Fig.~4 for $c_2=0.7$ (dashed curve). (The numerical factor 0.8 is again suitably chosen.) The good agreement of two plots in Fig.~6(b) and 6(c) implies that the competitive dynamics among dendritic branches starting from a linear seed and that for sidebranches in the region far from the dendritic tip are essentially the same.

The tip velocity of the largest branch is small at small $n$ and it increases with time as is seen in Fig.~6(a).  This is because the spacings between the survived branches increase with time and the tip velocities become larger for the larger spacing. To understand the relation between the tip velocity and the spacing between the branches,  
we have performed numerical simulations in a rectangular box of size $L_x\times L_y$ with periodic boundary conditions in the direction of $x$. As an initial condition, we have put a small branch of size $3\times 10$ at $i=L_x/2$ and $j=1$.  Since we have assumed periodic boundary conditions, the growth condition is equivalent to the case involving many branches, which have completely the same size and are periodically located with the same interval $L_x$, in an infinitely wide system. The system size $L_y$ is fixed to be 700, and $L_x$ is changed from 10 to 120, and the parameter $c_2$ is 0.7. Figure 7(a) displays the time evolution of the tip position of the branch in a logarithmic scale.  The time evolution seems to obey a power law $j_p=n^{\alpha}$ with exponent 0.7 in fairly wide region of $t$. 
In a narrow channel, the steady growth of dendrite is not possible when the undercooling is small. Brener et al. investigated the parameter region for the existence of steadily growing dendrites.\cite{rf:18} Figure 7(b) displays numerically evaluated exponents $\alpha$ as a function of $L_x$. The flat interface grows with a power law of exponent 1/2, which is shown  theoretically and was numerically shown  also in our coupled map lattice in the previous paper,\cite{rf:11} which corresponds to the limit of $L_x=0$. Inversely, 
 the exponent becomes 1, when the undercooling and the spacing $L_x$ are larger than a certain threshold, and the steady growth becomes possible.  Figure 7(b) shows that the exponent increases with the spacing $L_x$ for our kinetic dendrites.  

In the crystal growth from a linear seed, the interface is almost flat and the growth exponent is nearly 1/2 in an initial stage. Then, many small branches are created. At this stage, the branch spacing is small, the growth is slow, and the growth exponent is still close to 1/2. As the competitive process proceeds, the average spacing between active branches increases as a result of competition, and the tip velocities of the survived branches increase further. In this time scale, the tip position seems to grow like a power law with exponent larger than 1. The exponents of the power law may be evaluated roughly as follows. If the tip velocity $v_P$ grows in a power law with exponents $\alpha$  as shown in Fig.~7(b), the exponents depend on the average spacing $\bar{l}$, and the average spacing $\bar{l}$ increases roughly as $\bar{l}\sim t$, the tip position will grow as $j_P\sim \int v_P dt\sim \int d/dt\{t^{\alpha(\bar{l})}\} dt\sim \int dt\{\alpha t^{\alpha-1}+(d\alpha/dt)t^{\alpha}\ln t\}$. If $\alpha$ grows as $\alpha\sim 0.5+ct$ for small $t$ with a certain constant $c$, $j_P\sim t^{1.5}$ and the exponent for the tip position is evaluated as $1.5$ for small $t$, although this is a very rough approximation.  If the spacing becomes comparable or larger than the diffusion length, the tip velocities of survived branches approach the constant tip velocity for a single branch in an infinitely wide system.  

\section{Summary}
We have performed a large scale numerical simulation of kinetic dendrites with a coupled map lattice model. 
Well-developed sidebranches have appeared in the coupled map lattice. Many sidebranches are naturally created and the competitive dynamics among sidebranches determines the shape of the envelope.
The envelope has a parabolic shape near the tip of the dendrite, and the angle of the envelope with respect to the main stem grows up to 45$^{\circ}$ finally. In an intermediate region, the envelope grows, roughly speaking, in a power law, however, the exponent seems to increase gradually as a function of the distance from the tip. 

Similar competitive dynamics among many dendritic branches is observed in the unidirectional growth.  We could compare quantitatively the temporal evolution of the unidirectional growth with the envelope of sidebranches as a function of the distance from the tip position. The temporal evolution of a single branch in a narrow rectangular box seems to obey a power law in a wide range of $t$. The exponent increases with the width of the rectangular box.  The width can be interpreted to correspond to the spacing betweeen the sidebranches for the dendritic growth. As a result of the competition among sidebranches, the spacing between the active sidebranches increases, and then the tip velocities of the sidebranches  are accelerated further.  When the spacing becomes sufficiently large, the tip velocities of the active branches become nearly equal to the tip velocity of a single branch in an infinitely wide system.   

We have calculated numerically the spacing $d(x)$ and the envelope
 $y(x)$ of sidebranches. The competitive dynamics among the active sidebranches via the diffusion field determines $d(x)$ and $y(x)$. Expecting that the large scale competitive dynamics may not depend on the surface tension effect and we have neglected it in this paper. The surface tension is important to determine the tip radius and the tip velocity as in the experiments by Glicksman et al. It is not difficult to involve the surface tension effect in our model. However, the surface tension effect makes the parabolic shape of the dendrites stable, and strong noises are necessary to construct well-developed sidebranches. It is a future problem to study the sidebranch structure for the surface tension dendrites.

\begin{thebibliography}{99}
\bibitem{rf:1}
J.~S.~Langer: Rev. Mod. Phys.  {\bf 52} (1980) 1. 
\bibitem{rf:2}
T.~Vicsek: {\it Fractal Growth Phenomena} (World-Scientific, Singapore, 1991) 2nd ed.
\bibitem{rf:3}
S.~C.Huang and M.~E.~Glicksman: Acta Metal.  {\bf 29} (1981) 701.
\bibitem{rf:4}
H.~Honjo and Y.~Sawada: J. Cryst. Growth  {\bf 58} (1982) 297.
\bibitem{rf:5}
A.~Dougherty, P.~D.~Kaplan and J.~P.~Gollub: Phys. Rev. Lett.  {\bf 58} (1987) 1652.
\bibitem{rf:6}
R.~Pieters and J.~S.~Langer: Phys. Rev. Lett. {\bf 56} (1986) 1948.
\bibitem{rf:7} 
S.~C.Huang and M.~E.~Glicksman: Acta Metal. {\bf 29} (1981) 717.
\bibitem{rf:8}
Q.~Li and C.~Beckermann: Phys. Rev. E  {\bf 57} (1998) 3176.
\bibitem{rf:9}
T.~Honda and H.~Honjo: talk at the 58th annual meeting of the Physical Society of Japan (2003).
\bibitem{rf:10}
H.~Sakaguchi: J. Phys. Soc. Jpn.  {\bf 67} (1998) 96. 
\bibitem{rf:11}
H.~Sakaguchi and M.~Ohtaki: Physica A {\bf 272} (1999) 300.
\bibitem{rf:12}
A.~Karma and W.~J.~Rappel: Phys. Rev. E  {\bf 53} (1996) R3017. 
\bibitem{rf:13} K.~Kaneko: in {\it Formation, Dynamics and Statistics of Patterns}  (World Scientific, Singapole, 1990) ed. by K.~Kawasaki, M.~Suzuki and A.~Onuki.
\bibitem{rf:14} D.~A.~Kessler, H.~Levine and W.~N.~Reynolds: Phys. Rev. A {\bf 42} (1990) 6125.
\bibitem{rf:15} N.~H.~Packard: in {\it Theory and Application of Cellular Automata} (World Scientific, 1986) ed. by S.~Wolfram.
\bibitem{rf:16} F.~Liu and N.~Goldenfeld: Physica D {\bf 47} (1991) 124.
\bibitem{rf:17}
E.~Ben-Jacob, P.~Garik, T.~Mueller and D.~Grier: Phys. Rev. A {\bf 38} (1988) 1370.
\bibitem{rf:18}
E.~A.~Brener, M.~B.~Geilikman and D.~E.~Temkin: Sov. Phys. JETP {\bf 67} (1988) 1002.
\end{thebibliography}
\end{document}